# Vision-Language Models Reproduce a Human Mental-Extrapolation Bias in the Watanabe Illusion

**Eiji Watanabe**[1,2,3,*], **Kyohei Ueda**[1,2], **Kagayaki Kato**[2,4], **Masaki Watabe**[1,5], **Akiyoshi Kitaoka**[6]

[1] National Institute for Basic Biology, Higashiyama 5-1, Myodaiji, Okazaki, Aichi 444-8787, Japan
[2] Department of Basic Biology, The Graduate University for Advanced Studies (SOKENDAI), Shonan Village, Hayama, Kanagawa 240-0193, Japan
[3] AI Facility, Trans-Scale Biology Center, National Institute for Basic Biology, Higashiyama 5-1, Myodaiji, Okazaki, Aichi 444-8787, Japan
[4] Optics and Imaging Facility, Trans-Scale Biology Center, National Institute for Basic Biology, National Institutes of Natural Sciences, Nishigonaka 38, Myodaiji, Okazaki, Aichi 444-8585, Japan
[5] Exploratory Research Center on Life and Living Systems (ExCELLS), Higashiyama 5-1, Myodaiji, Okazaki, Aichi 444-8787, Japan
[6] College of Comprehensive Psychology, Ritsumeikan University, Ibaraki, Osaka, Japan

[*] For correspondence: eiji@nibb.ac.jp

## Abstract

Vision-language models (VLMs) are increasingly used for visual judgments, yet whether they share human perceptual biases, and where in the model such biases arise, is unclear. We addressed this with the Watanabe Illusion, a stimulus not previously studied quantitatively: an oblique segment inside a circle whose extension geometrically hits the topmost of 11 dots, yet human observers (n = 131) reported a dot near the middle of the column. Five VLMs (Qwen2.5-VL-32B and 7B, Claude Sonnet 4.6, GPT-5.5 Instant, Gemini 3.5 Flash), given the identical stimulus and wording, all deviated in the same direction. In the open-weight Qwen2.5-VL-32B, linear probes on the vision encoder recovered the line angle and the target dot six to eight times more accurately than its behavioral responses. A layer-wise logit lens showed that the correct answer dominates in the middle layers of the language model and is overwritten in late layers only under a visual-judgment frame. Naming the illusion, requesting suppression of prior knowledge, forcing a visual chain of thought, and 390 semantic preambles left the response center unchanged; only disclosing the correct answer moved it. VLMs may therefore carry human perceptual biases at the language-generation stage, where they resist declarative correction.

# Impact statement

Vision-language models encode a line's true direction accurately yet report the same mis-extrapolated target as human observers, a bias imposed in late language layers and unaffected by prompting.

# Introduction

Vision-language models (VLMs), large models that take images as input and respond in natural language, have proliferated rapidly and are beginning to be used routinely for visual-judgment tasks, including medical imaging, technical-drawing interpretation, and measurement support. Yet little is known about how these models "see" a stimulus, and in particular whether they share the biases specific to human perception.

Visual illusions are a classical tool for exposing the dissociation between perception and physical reality, and susceptibility to them has become one criterion for gauging how human-like a system's vision is. For instance, a growing body of research evaluates artificial neural networks as models of human vision by testing whether they reproduce illusory percepts (e.g., Watanabe et al., 2018; Lotter et al., 2020; Ueda et al., 2026). Studies presenting illusion stimuli to VLMs have also emerged in recent years; benchmarks spanning multiple classical illusions have reported that VLM responses partially agree with human perception, but that it is difficult to determine whether this agreement reflects perceptual understanding or a surface pattern derived from the training data (Shahgir et al., 2024; Zhang et al., 2023). In particular, Ullman (2024) showed that VLMs erroneously judge stimuli that resemble illusions but produce no illusory effect to be "illusions," raising the possibility that VLM responses to illusions reflect the recollection of illusion commentary contained in the training data. Two gaps remain in the existing literature. First, because classical illusions are accompanied by countless explanations and human response data on the internet, it is difficult to determine whether a VLM response is “the result of looking at the image” or “the result of recalling a description of the illusion.” Second, few studies have localized the origin of illusion responses through direct analysis of the model’s internal representations.

This paper addresses both gaps by examining a single stimulus in depth. The stimulus is the Watanabe Illusion (Figure 1a), which one of the authors has made public on Figshare (Watanabe, 2018). As a related phenomenon, Bouma and Andriessen (1968) reported that the orientation of line segments close to horizontal or vertical is perceived as closer to horizontal or vertical, and showed a similar bias in judgments of collinearity with a distant dot. That study did not, however, treat the effect as an illusion, and it has since been cited mainly in the context of tilt normalization and the oblique effect. In a short phenomenological report that re-evaluates this phenomenon as an illusion, Usui and Kitaoka (2025) depicted the present stimulus as a structural variant. No quantitative study has addressed the stimulus itself, and the present paper is its first detailed report. These works are rarely cited, and Usui and Kitaoka (2025) in particular was published after the release of Qwen2.5-VL-32B (March 2025), the model used here for internal analysis. Contamination through the training data is therefore expected to be limited. None of the four VLMs we tested (Qwen2.5-VL-32B, Claude Opus 4.7, GPT-5.5, Gemini 3.5 Flash) could name this stimulus as a known illusion, whereas the three commercial frontier models named the classical illusions (Müller-Lyer, Poggendorff, Kanizsa) accurately (Appendix 1).

We therefore used the Watanabe Illusion as the test stimulus and carried out (i) a survey of the illusion in human observers (n = 131), (ii) a behavioral survey of five VLMs under an identical protocol, (iii) localization of the origin of the bias through internal analysis of the open-weight model Qwen2.5-VL-32B, and (iv) tests of whether declarative interventions and 390 semantic interventions can shift the bias.

# Results

## Human observers show a mental-extrapolation bias in the Watanabe Illusion

The stimulus (104.jpg, Figure 1a) contains, inside a circle at the lower-left corner, a line segment inclined 23.5° from the horizontal; 11 equally spaced dots run down the right edge. The geometric extension of the segment intersects the topmost dot (dot 1), but no extension line or guide is drawn, so the observer must extrapolate the intersection from the apparent trajectory of the segment alone. In an online experiment, participants answered two independent questions asking which dot the line would hit, counted from the top ($Q_1$) and from the bottom ($Q_2$). Illusion magnitude was summarized by the Illusion Index ($II$), which is 0 for a veridical response and 10 for the maximal illusion (Materials and methods).

Summaries of human $Q_1$, $Q_2$, and $II$ are shown in Figure 1c–e (Human), with detailed statistics in Appendix 2—table 1. Although the line physically passes through dot 1, human observers (n = 131) on average reported the 4.65th dot from the top ($Q_1$, SD = 2.63) and the 6.36th dot from the bottom ($Q_2$, SD = 2.86). The mean $II$ was 4.15 (SD = 2.28). Only 4.6% of participants (6 of 131) gave a veridical response ($Q_1 = 1$). We call this systematic downward deviation the mental-extrapolation bias. It agrees with the phenomenological observation of Usui and Kitaoka (2025) for this stimulus (observers tend to point to the third or fourth dot from the top), and it is in the same direction as the report of Bouma and Andriessen (1968) that near-horizontal line segments are perceived as closer to horizontal. At the group level the answers were internally consistent: an observer identifying a single dot should produce $Q_1 + Q_2 = 12$, and the human mean was 11.01 (SD = 3.07). They were also asymmetric, with $IIQ_2$ (4.64) exceeding $IIQ_1$ (3.65) by about one dot. The recognition probe in Appendix 1 confirms that this stimulus is rare within the training distribution of contemporary VLMs (0 correct namings in 80 naming trials across four VLMs, against up to 100% for the classical-illusion controls).

## VLMs reproduce the same bias as humans

Figure 1b shows the phenomenon in a single image. We passed 104.jpg to the image-generation function of ChatGPT (GPT-5.5) and asked only that it add a star to the dot the line segment hits. In the output image, although the line segment was nearly preserved (23.3°), the star was placed near dot 6, within the range of the behavioral bias of humans and VLMs. Anyone can replicate this demonstration in a few minutes. We then quantified the phenomenon under a controlled protocol in which five VLMs received the same wording and the same independent-trial structure as the human participants (Materials and methods).

Figure 1c–e shows the mean ± SD of $Q_1$, $Q_2$, and $II$ for all six data sources (humans and VLMs). The group means of $II$ were 4.91 for Qwen-32B, 4.95 for Qwen-7B, 5.17 for Sonnet 4.6, 6.04 for Gemini 3.5 Flash, and 7.78 for GPT-5.5 Instant. All five deviated in the same direction as humans (4.15) and lay far from veridical ($II$ = 0): each model reported a dot in the middle-to-lower range rather than dot 1.

Veridical responses ($Q_1 = 1$) occurred in only 2 of the 299 trials pooled across the five VLMs (both from Qwen-7B). The directional asymmetry observed in humans ($I\,I\,Q_2 > I\,I\,Q_1$) was also reproduced at both Qwen sizes (32B: +1.38; 7B: +2.36; humans: +0.99).

Qwen-32B compressed the spread of its responses to about one quarter of human inter-individual variability ($I\,I$ SD = 0.63, 0.28× the human value), whereas the 7B model of the same series and architecture retained variance close to that of humans ($Q_1$ SD = 2.51, 0.95× the human value of 2.63; $I\,I$ SD = 1.75). The variance of the three commercial models lay between the two ($I\,I$ SD = 0.75–1.29). How much a model compresses response variance therefore differs greatly between models, even within one model series. All statistics are collected in Appendix 2—table 2.

## The vision encoder retains accurate information about the line's inclination

To test whether visual encoding preserves the angle of the segment, we directly probed the vision encoder of the same model whose behavior we had measured (Qwen2.5-VL-32B), using linear probes trained on a parametric stimulus set of 68 images (Materials and methods).

The line angle was recovered with $R^2$ = 0.985 and MAE = 2.23° from the final ViT block (Representation A; Figure 2a), and performance was comparable at the Patch Merger output that the language model actually receives ($R^2$ = 0.981, MAE = 2.43°). The dot number indicated by the line segment was recovered with sub-dot accuracy, MAE = 0.61 dots (Representation C; Figure 2b). Stimulus condition (presence of the circle, length of the segment) had little effect on probe accuracy.

We compared this internal accuracy with the behavior of the same model (Appendix 2—table 3). The behavioral error was 4.22 dots for $Q_1$ and 5.60 dots for $Q_2$, roughly six to eight times the linear read-out error from the encoder (0.71 dots). The encoder therefore passes accurate information to the language model, and the bias is most likely introduced downstream of that interface.

## The bias arises in the late layers of the language model

To localize where within the language model the bias arises, we applied a layer-wise logit lens (Belrose et al., 2023; nostalgebraist, 2020) to the 64 layers of Qwen2.5-VL-32B, prefilling the assistant response with one of four fixed continuations and reading out the probabilities of the digit tokens “1”–“9” at each layer (n = 10 trials per continuation; Materials and methods):

- **v1** “The line would hit the” (visual-judgment baseline)
- **v2** “The topmost dot is the” (concept-recall control)
- **v3** “The number five is the” (irrelevant-prior control)
- **v4** “The line would hit dot number” (visual judgment + numerical cue)

Figure 3 shows the layer × digit-probability curves for the four conditions. In v1, v2, and v4 alike, the correct digit “1” rises steeply in the middle layers (layers 35–45) and briefly becomes the argmax (peak $P \approx 0.40$–$0.50$). The model thus appears to form the correct candidate partway through processing, and what happens to it afterwards depends on the continuation. Under the concept-recall frame (v2), “1” persists to the final layer ($P = 0.94$). Under the visual-judgment frames (v1, v4), “1” is systematically overwritten in the late layers (~45–63), and the final layer commits to a distribution centered on “5.” In v3, “5” is carried from the middle layers onward and persists. This suggests that a context-dependent override mechanism in the late layers of the language model, triggered when the input is framed as a visual judgment, rewrites the correct middle-layer candidate into a response consistent with human reports. Final-layer values are given in Appendix 2—table 4.

## The bias resists declarative and semantic interventions

We tested whether the late-layer override could be disabled by declarative linguistic interventions, using an eight-condition prompt-cancellation experiment (n = 20 each, $Q_1$; Figure 4). Naming the illusion (P1: “Watanabe Illusion”; P2: “a visual illusion”), requesting the suppression of prior knowledge (P3: “Ignore any prior knowledge… answer based only on the line’s pixel-level geometry”), and forcing a visual chain of thought (P5) all failed to move the response outside the baseline band (P0 mean 5.45 ± 1 SD). An irrelevant context (P7) behaved likewise. The only manipulations that moved the response were direct content interventions: a counter-hint (P4) produced an incomplete shift of −1.25 dots, and only disclosure of the correct answer (P6) collapsed the response to near-veridical (mean 1.35) (condition-wise statistics in Appendix 2—table 5). The full text of every prompt is recorded in Appendix 3.

Although the model is fully capable of generating the words “ignore prior knowledge,” that request does not reach the process that produces the bias. The bias does not reside at the level of declarative knowledge that the language model can introspect on, name, and rewrite. Semantic interventions drawn from a much broader space did not shift it either. In nine conditions ($n \approx 130$ each, one-shot design) in which 390 preambles describing the observer's identity (a sniper, a poet, an octopus), living environment (dense fog, zero gravity), or a neutral situation were placed in the system or user message, the response mean remained within the narrow range 5.52–6.40, and no preamble moved the response center toward veridical (Appendix 2—table 6). Placing the preamble in the system message increased response variance (maximum SD = 1.75, 67% of the human inter-individual SD of 2.63), but the center did not move, and no condition reached the human inter-individual variance.

# Discussion

## What the results suggest

All five VLMs showed a bias in the same direction as humans, of comparable or greater magnitude. Because the illusion is rare in the VLMs' training distribution, these responses are unlikely to be a memory of a particular illusion (Appendix 1). In Qwen2.5-VL-32B, the vision encoder retained accurate angle information, and the bias arose through a context-dependent override in the late layers of the language model. Neither declarative interventions nor 390 semantic interventions moved the response center.

Although the vision encoder retains accurate information, a VLM trained on human-derived data (human descriptions of and annotations on images) may hold human perceptual biases at the language-generation stage, where they are deeply embedded and hard to remove. The training labels of a VLM are thought to reflect not pixel-level geometry but what human observers "report having seen." The fact that all VLMs showed a bias in the same direction despite the Watanabe Illusion being a rare stimulus suggests that this "knowledge" is not a memory of a particular illusion but a distillation of the statistical structure of human perceptual responses. The tilt normalization shown by Bouma and Andriessen (1968) is a general property of human perception that is not restricted to particular figures. Even when an individual stimulus is unknown, if such general properties are reflected in the training data through human descriptions, a VLM can show a bias in the same direction as humans for that unknown stimulus.

## Application risks and contributions to basic science

VLMs are being introduced into practical image analysis in medicine, manufacturing, and measurement. Our results show that even when accurate information exists inside the model, the linguistic response can be systematically distorted by a human-like bias, and that this bias cannot be removed by prompt-level warnings or requests for suppression. When VLM visual judgments are used in practice, this class of bias needs to be investigated comprehensively, beyond illusion stimuli.

If VLMs distill human report statistics, they may also be useful as models of human vision. The dissociation observed here (accurate information exists internally, yet the report is biased; and the bias is not under the control of declarative knowledge) parallels the human experience of “seeing an illusion even while knowing it is an illusion.” Because the layer where this transformation occurs can be located in a model, the same question can be put to humans: do early measurements that precede verbal report carry veridical information?

## Limitations

(i) The behavioral comparison rests on a single stimulus (104.jpg, 23.5°), and the angle dependence of the bias has not been tested. Future experiments should use stimuli in which the physically correct dot lies in the middle of the column. (ii) The internal analysis is limited to the open-weight Qwen2.5-VL-32B, and whether the same mechanism operates inside commercial VLMs can only be inferred from the agreement in behavioral patterns. (iii) The values for commercial VLMs are a snapshot of default web-UI behavior as of May 2026. (iv) The success of a linear probe establishes the presence of information, but not its absence. (v) The variance value for 7B is based on a single $n = 60$ run, and confirmation of run-to-run stability would be desirable. (vi) The human $n = 131$ reflects between-subject variability, whereas the VLM variance derives from stochastic decoding of a single model, so this asymmetry remains in any comparison of the two variances. (vii) Because Usui and Kitaoka (2025) depicts this stimulus, we cannot rule out that it is included in the training data of commercial models trained after that paper’s publication. In the recognition probe of Appendix 1, however, none of the models could name the stimulus.

# Materials and methods

## Key resources table

| Reagent type (species) or resource | Designation | Source or reference | Identifiers | Additional information |
|---|---|---|---|---|
| Software, algorithm | Qwen2.5-VL-32B-Instruct | Bai et al. (2025) | Hugging Face: Qwen/Qwen2.5-VL-32B-Instruct | Open weights, BF16 |
| Software, algorithm | Qwen2.5-VL-7B-Instruct | Bai et al. (2025) | Hugging Face: Qwen/Qwen2.5-VL-7B-Instruct | Open weights, BF16 |
| Software, algorithm | Claude Sonnet 4.6; Claude Opus 4.7 | Anthropic | — | Web UI (behavior); API (recognition probe) |
| Software, algorithm | GPT-5.5 Instant; GPT-5.5 | OpenAI | — | Web UI (behavior); API (recognition probe) |
| Software, algorithm | Gemini 3.5 Flash | Google | — | Web UI (behavior); API (recognition probe) |
| Software, algorithm | Gorilla Experiment Builder | Gorilla | https://app.gorilla.sc/openmaterials/1339686 (demonstratio | Online human experiment |

| Reagent type (species) or resource | Designation | Source or reference | Identifiers | Additional information |
|---|---|---|---|---|
| | | | n) | |
| Software, algorithm | Analysis and experiment code | This paper | GitHub: https://github.com/eijwat/watanabe-illusion-vlm; Zenodo: doi:10.5281/zenodo.23096369 | See Appendix 4 |
| Other | Watanabe Illusion stimulus (104.jpg) | Watanabe (2018) | doi:10.6084/m9.figshare.5838606 | 539 × 264 px, 23.5° |

## Stimulus

The stimulus (104.jpg, Figure 1a) is a rectangle (539 × 264 px). Inside a circle at the lower-left corner, a line segment inclined 23.5° from the horizontal is drawn, and 11 equally spaced dots are arranged from top to bottom along the right edge. The segment is designed so that its geometric extension intersects the topmost dot (dot 1). No extension line or guide is drawn; the observer must extrapolate the intersection solely from the apparent trajectory of the segment.

## Human experiment

In an online psychophysical experiment (Gorilla platform), participants were presented with 104.jpg and asked to answer the following two questions independently.

- $Q_1$ **(from top):** *Estimate which dot on the right edge the line inside the circle would hit if extended. Answer like 'the Xth dot from the top.'*
- $Q_2$ **(from bottom):** *Estimate which dot on the right edge the line inside the circle would hit if extended. Answer like 'the Xth dot from the bottom.'*

After the exclusions described under Participants, restricting the analysis to participants who answered both $Q_1$ and $Q_2$ with values within [1, 11] yielded data from 131 participants. The experiment as run in Gorilla can be tried in an online demonstration (https://app.gorilla.sc/openmaterials/1339686).

**Participants.** Participants were recruited through Prolific (https://www.prolific.com/). The main experiment was run on 23 April 2026 with 198 participants before exclusion; a pilot test with 94 participants was run on 16 April 2026, and its data were not used. Participants were paid at an hourly rate of £12.00. Before analysis, participants were excluded if they failed the attention check, if their answer to an additional question on the angle of the line (not analyzed in this study) was invalid, or if they had participated more than once. Demographic data from the Prolific profiles were available for the 195 participants whose submissions were approved on Prolific: self-reported sex was 50.3% male (n = 98), 49.2% female (n = 96), and 0.5% not reported (n = 1), and age was 33.38 ± 10.25 years (mean ± SD; male 33.37 ± 10.21, female 33.50 ± 10.34).

**Ethics.** The experimental procedure received approval from the Life Science Research Ethics Committee of the National Institutes of Natural Sciences (Project ID: EC01-078), and informed consent was obtained from all participants. All methods were carried out in accordance with relevant guidelines and regulations.

## Illusion Index

Illusion magnitude is expressed by the Illusion Index ($II$):

$$II = \frac{(Q_1 - 1) + (11 - Q_2)}{2}$$

A veridical response ($Q_1$ = 1, $Q_2$ = 11) gives $II$ = 0, and the maximal illusion ($Q_1$ = 11, $Q_2$ = 1) gives $II$ = 10. The direction-specific sub-indices are $IIQ_1 = Q_1 - 1$ and $IIQ_2 = 11 - Q_2$.

## VLM behavioral experiment

We presented the VLMs with the same wording and the same independent-trial structure as the human experiment ($Q_1$ and $Q_2$ asked in separate turns with no conversation history). We tested five VLMs:

- **Open weights (n = 60 each):** Qwen2.5-VL-32B and 7B (Bai et al., 2025) (BF16, temperature = 1.0, fixed per-trial seeds). Only the model ID differed between the two runs (same script and prompt), enabling a comparison across model sizes. For both sizes, we confirmed via image reconstruction from the processor output that the stimulus was received at native dynamic resolution without distortion.

- **Commercial frontier models (web UI, default mode, n = 60 each):** Claude Sonnet 4.6, GPT-5.5 Instant, and Gemini 3.5 Flash. Each trial was conducted in a new temporary chat with memory storage turned off. We used the web UI because it reflects "the VLM that people actually experience." Data were collected in May 2026.

The image-generation demonstration (Figure 1b) passed 104.jpg to the image-generation function of ChatGPT (GPT-5.5) with a request only to add a star to the dot the line segment hits.

## Statistics

All six data sources were processed with the same response parser, the same directional normalization (responses given in the opposite counting direction were mapped by value → 11 − value + 1), and the sample standard deviation (ddof = 1). Because a response could not be extracted on some trials, the Illusion Index and the sum of the two answers were computed from the trials in which both answers were extracted, pairing them by trial (paired n = 58 for Qwen-7B and 59 for Gemini). Results are reported descriptively as mean ± SD (with medians and ranges in Appendix 2); no inferential tests were applied.

## Vision encoder probes

We generated a parametric stimulus set (17 angles × 4 conditions = 68 images, generate_stimuli.py) and trained linear probes with ridge regression under leave-one-out cross-validation on the final ViT block output (Representation A) and the Patch Merger output (Representations B and C) of Qwen2.5-VL-32B. Targets were the line angle and the dot number indicated by the line. Probe accuracy is reported as $R^2$ and mean absolute error (MAE).

## Logit lens

For each of the four continuations (v1–v4, Results), the assistant response was prefilled with the continuation and, at the final token position, the within-layer softmax probabilities of the digit tokens "1" through "9" were read out across all 64 layers of the language model of Qwen2.5-VL-32B (Belrose et al., 2023; nostalgebraist, 2020); n = 10 trials per continuation.

## Prompt cancellation

Eight conditions (P0–P7) × n = 20 = 160 single-turn trials on Qwen2.5-VL-32B-Instruct, $Q_1$ only, no conversation history, seed = (prompt index) × 1000 + trial index. Each prompt was the fixed preamble followed by the condition-specific body (Appendix 3).

## Semantic preamble intervention

A set of 390 preambles (130 in each of three categories: Character, Environment, and Neutral) was placed in either the system or the user message in nine conditions (A–I) on Qwen2.5-VL-32B-Instruct, $Q_1$ only. The design is one-shot: each condition comprises 130 trials, each with a different preamble and a different seed, and no preamble is repeated. Condition definitions, seed scheme, and message structure are given in Appendix 3.

## Recognition probe

Four stimuli (104.jpg and three classical illusions) were presented to Qwen2.5-VL-32B (A100) and, via official APIs without conversation history, to Claude Opus 4.7, GPT-5.5, and Gemini 3.5 Flash, with three question types in independent trials (n = 10 per cell, 480 trials in total). Details are given in Appendix 1.

## Generation settings and computing environment

Open-weight runs used temperature = 1.0, top_p = 1.0, repetition_penalty = 1.05, max_new_tokens = 512, and a fixed per-trial seed, on an NVIDIA A100 80GB (x86_64) and an NVIDIA DGX Spark (GB10 Grace Blackwell, 128 GB unified memory, ARM64, CUDA 13.0). Full details are given in Appendix 4.

## Use of generative AI

The experimental design, analysis code, and documentation were developed in collaboration with Claude (Anthropic). ChatGPT (OpenAI) also played a substantial role in generating ideas. Both Claude and ChatGPT additionally participated in this study as objects of experimentation. The human authors take full responsibility for the scientific content, the integrity of the data, the interpretations drawn from the results, and any remaining errors.

# Additional information

## Funding

| Funder | Grant reference number | Author |
| --- | --- | --- |
| Japan Society for the Promotion of Science (JSPS) | Grants-in-Aid for Scientific Research (A, B, and C) | Eiji Watanabe |
| Japan Society for the Promotion of Science (JSPS) | KAKENHI JP21H04426 | Akiyoshi Kitaoka |
| Japan Science and Technology Agency (JST) | SPRING JPMJSP2104 | Kyohei Ueda |

The funders had no role in study design, data collection and interpretation, or the decision to submit the work for publication.

## Competing interests

The authors declare that no competing interests exist.

## Author contributions

Eiji Watanabe: Conceptualization, Methodology, Software, Validation, Formal analysis, Investigation, Resources, Data curation, Writing – original draft, Writing – review & editing, Visualization, Project administration, Funding acquisition
Kyohei Ueda: Methodology, Validation, Formal analysis, Investigation, Data curation, Funding acquisition
Kagayaki Kato: Software, Validation, Resources
Masaki Watabe: Software, Validation, Writing – original draft, Writing – review & editing, Visualization
Akiyoshi Kitaoka: Conceptualization, Validation, Resources, Supervision, Funding acquisition

## Ethics

Human subjects: the experimental procedure received approval from the Life Science Research Ethics Committee of the National Institutes of Natural Sciences (Project ID: EC01-078), and informed consent was obtained from all participants.

## Data availability

All code, stimuli, cleaned human response data, and raw model outputs are available at GitHub (https://github.com/eijwat/watanabe-illusion-vlm) and archived at Zenodo (https://doi.org/10.5281/zenodo.23096369); Appendix 4 maps every result in this paper to the files that reproduce it. The raw Gorilla exports of the human experiment are not released because they contain participant-identifying information; the cleaned responses used for all analyses (data/human/human_2nd_test.csv) are released. The main stimulus is available on Figshare (doi:10.6084/m9.figshare.5838606).

The following dataset was generated:

Watanabe E, Ueda K, Kato K, Watabe M, Kitaoka A. 2026. watanabe-illusion-vlm: code, stimuli and data for "Vision-Language Models Reproduce a Human Mental-Extrapolation Bias in the Watanabe Illusion". Zenodo. doi:10.5281/zenodo.23096369

# Figures

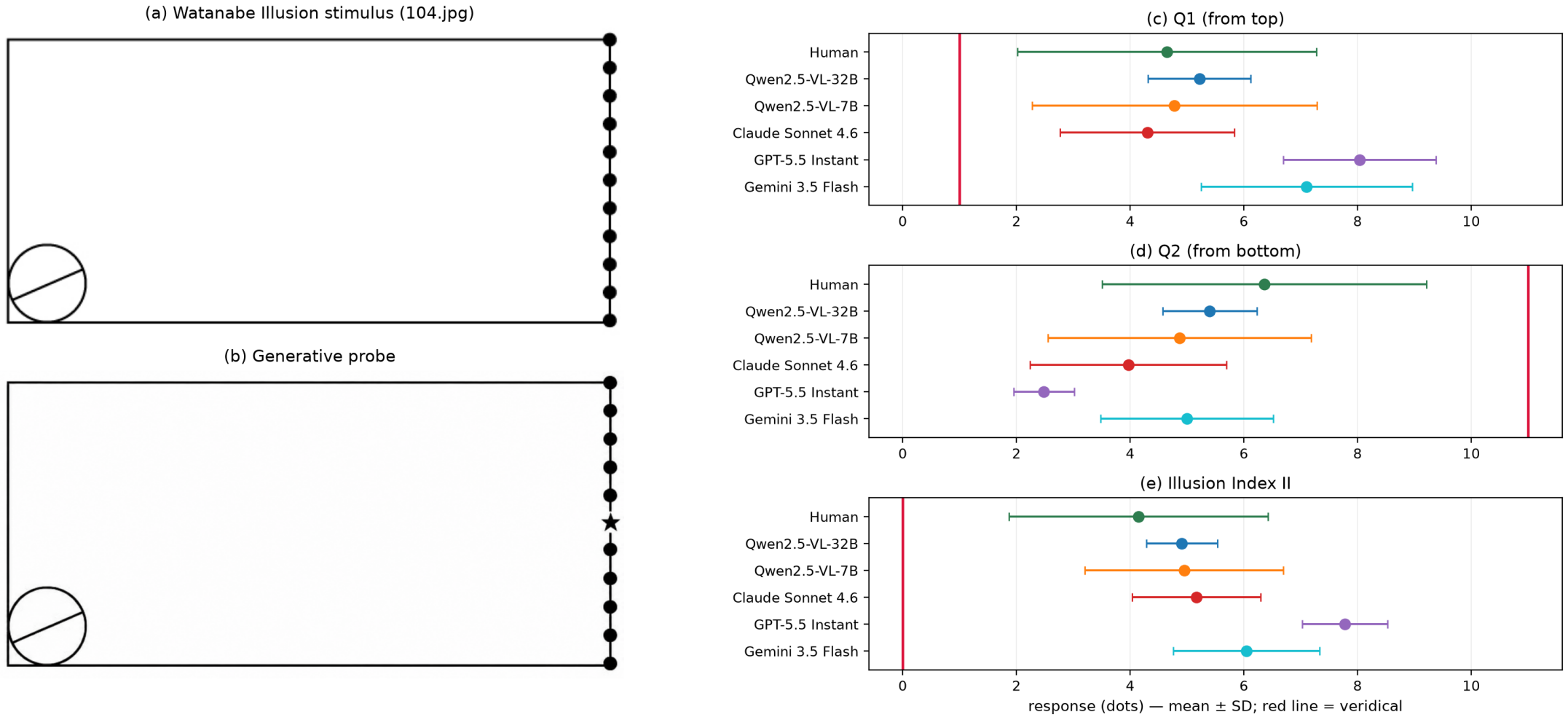

**Figure 1. The Watanabe Illusion and the responses of human and VLM observers.** (**a**) Main stimulus 104.jpg: the line segment, partially visible inside the circle, geometrically passes through the topmost dot (dot 1) of the column of 11 dots along the right edge. (**b**) Demonstration with a generative probe: the output when ChatGPT was asked only to add a star to the dot the line segment hits. Although the line segment is preserved at 23.3°, the star was placed near dot 6. (**c**–**e**) $Q_1$ (**c**), $Q_2$ (**d**), and the Illusion Index $II$ (**e**) for six data sources: humans (n = 131 participants), Qwen2.5-VL-32B (n = 60 trials), Qwen2.5-VL-7B (n = 60; paired n = 58), Claude Sonnet 4.6 (n = 60), GPT-5.5 Instant (n = 60), and Gemini 3.5 Flash (n = 60; paired n = 59). Points are means, bars are ± SD (sample SD, ddof = 1), and the red line marks veridical ($Q_1$ = 1, $Q_2$ = 11, $II$ = 0). All VLMs deviate from veridical in the same direction as humans. Statistics are descriptive; full values are in Appendix 2—table 2.

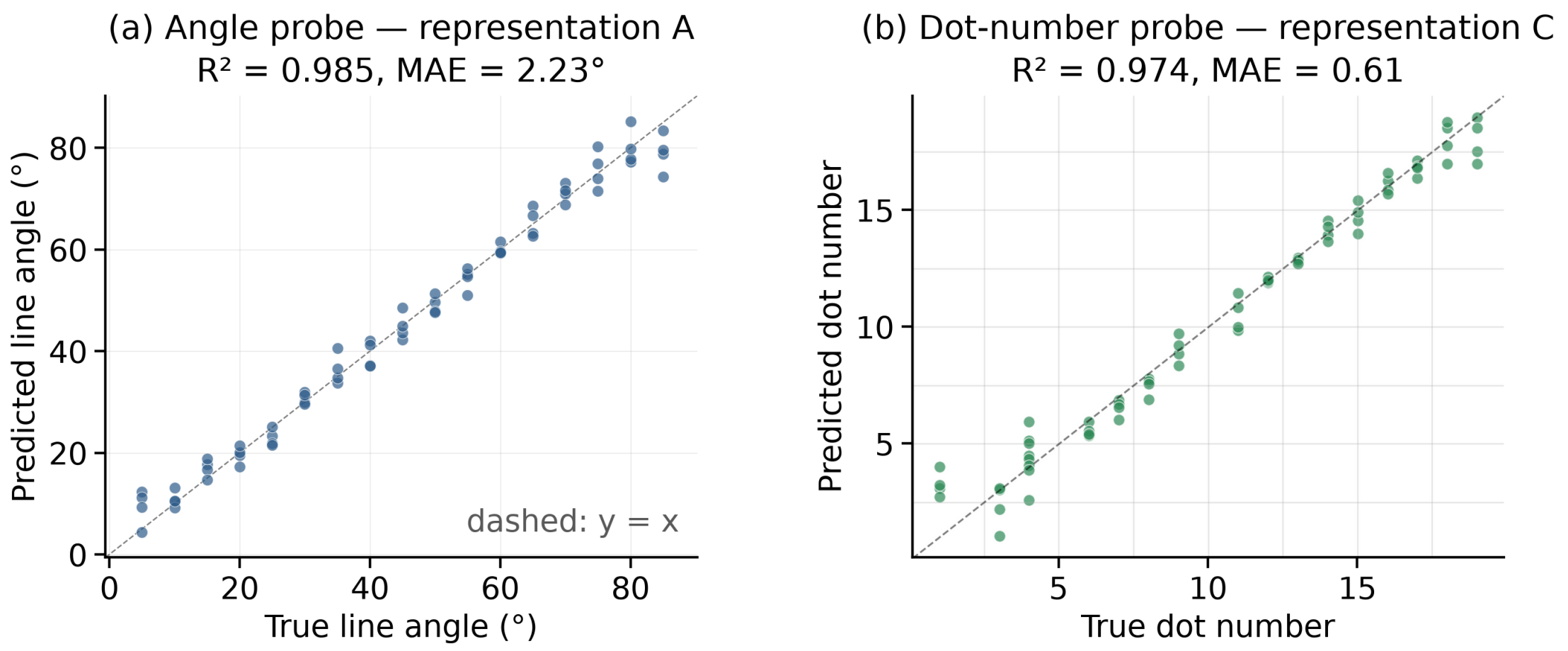

**Figure 2. Linear probes on the Qwen2.5-VL-32B vision encoder recover the stimulus geometry accurately.** Parametric stimulus set (n = 68 images), ridge regression with leave-one-out cross-validation. (**a**) Angle probe (Representation A = final ViT block): recovers the true line angle with $R^2$ = 0.985, MAE = 2.23°. (**b**) Dot-number probe (Representation C = Patch Merger flattened): recovers the dot number indicated by the line with $R^2$ = 0.974, MAE = 0.61 dots. This is roughly six to eight times more accurate than the behavioral error of the same model (4.22–5.60 dots; Appendix 2—table 3), showing that the bias is not present at the visual-encoding stage.

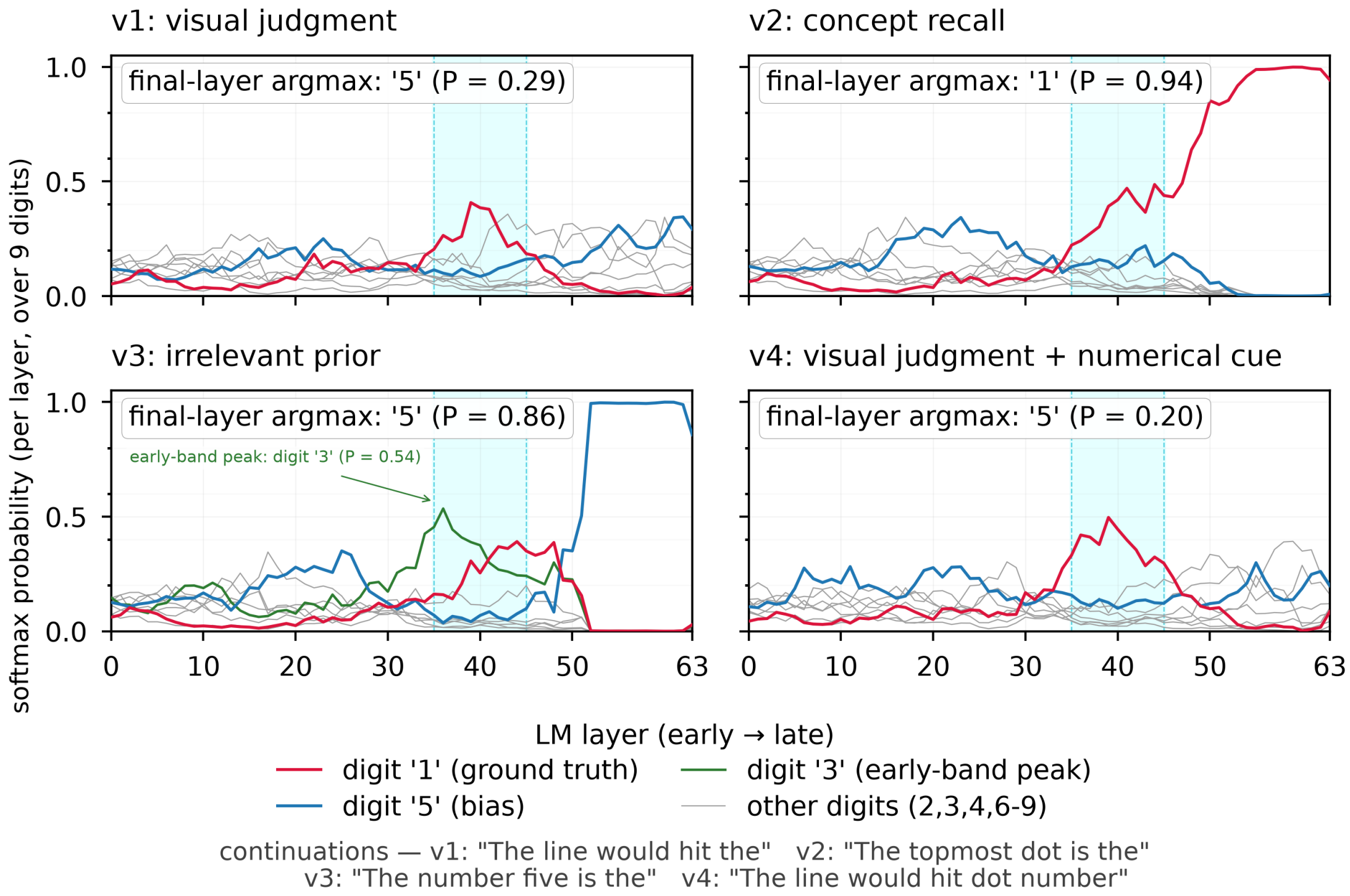

**Figure 3. Layer-wise logit lens across the 64 layers of the Qwen2.5-VL-32B language model.** Each panel shows, for one continuation (v1–v4), the within-layer softmax probabilities of digit tokens 1–9 at each layer (mean of n = 10 trials per continuation). The correct answer “1” is the bold red line and the canonical bias “5” is the bold blue line. In v1, v2, and v4, “1” rises to argmax dominance in the middle layers (~35–45). Under the concept-recall frame (v2), “1” persists to the final layer ($P$ = 0.94), whereas under the visual-judgment frames (v1, v4) it is overwritten in the late layers (~45–63) and the final layer commits to a distribution centered on “5.” In v3, “5” is carried from the middle layers onward (control). Final-layer values are in Appendix 2—table 4.

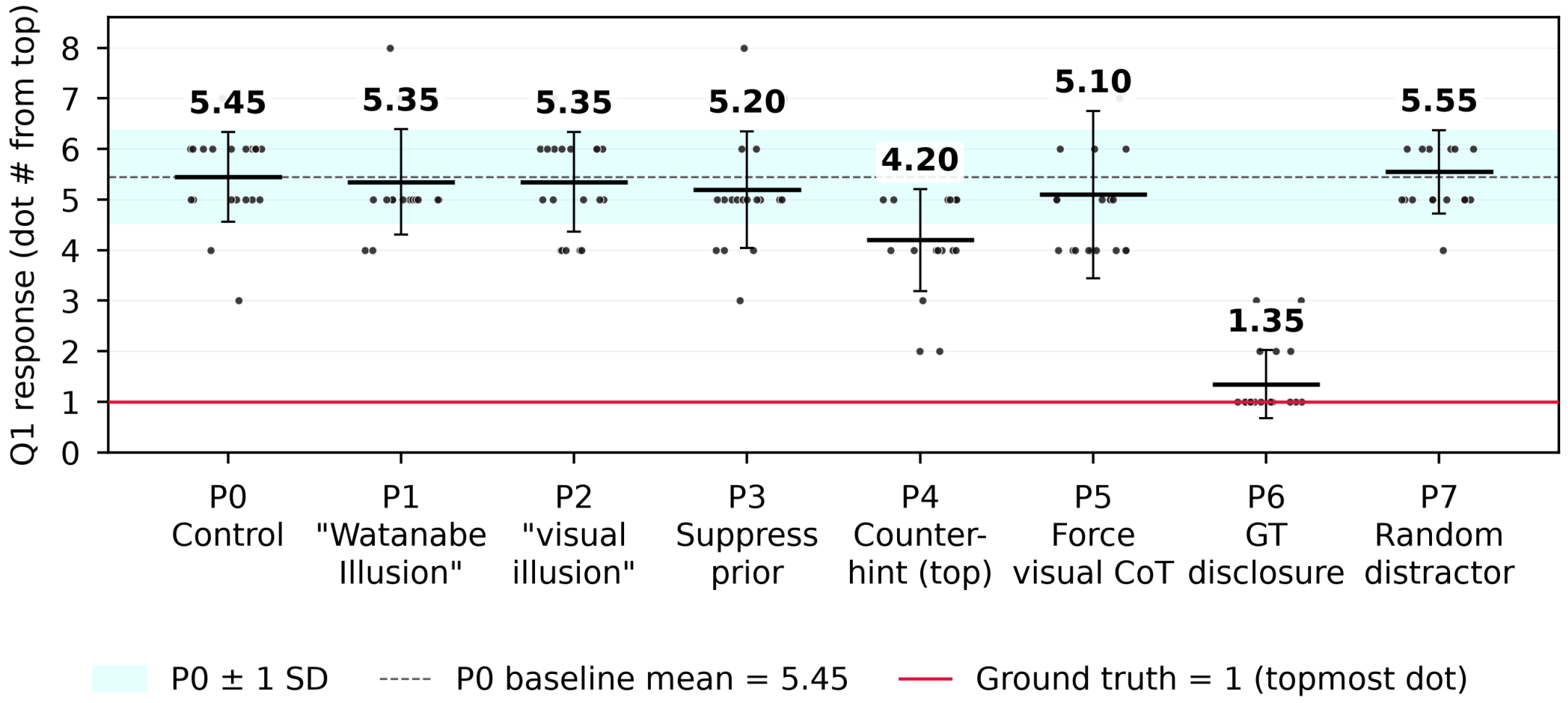

**Figure 4. Declarative prompt interventions do not move the biased response.** Mean $Q_1$ responses of Qwen2.5-VL-32B across the eight prompt-cancellation conditions (n = 20 trials each). Bars are mean ± SD, black points are individual trials, the dashed line is the P0 control mean (5.45), the shaded band is P0 ± 1 SD, and the red line marks veridical ($Q_1$ = 1). Naming the illusion (P1, P2), requesting the suppression of prior knowledge (P3), forcing a visual chain of thought (P5), and an irrelevant context (P7) all fail to move the response outside the P0 ± 1 SD band. The counter-hint (P4) produces an incomplete shift (−1.25 dots), and only disclosure of the correct answer (P6) collapses the response to near-veridical. Statistics are descriptive; condition-wise values are in Appendix 2—table 5, and the prompts in Appendix 3.

# Appendix

# Appendix 1. Recognition probe: the Watanabe Illusion is an unknown stimulus for contemporary VLMs

This appendix tests whether VLM responses are generated “by looking at the image content” or “retrieved from descriptions of the Watanabe Illusion in the training data.” To address it, we measured whether four VLMs could identify 104.jpg as the Watanabe Illusion, using three classical illusions as positive controls.

**Protocol.** Four stimuli (104.jpg, Müller-Lyer, Poggendorff, Kanizsa; the three classical stimuli are shown in Appendix 1—figure 1) were presented to each VLM, and three types of question were posed in independent trials: (QV1) yes/no familiarity, (QV2) forced naming of a known illusion, and (QV3) description + recognition. Search was prohibited (“Do not search the web.”). The models tested were Qwen2.5-VL-32B (A100), Claude Opus 4.7, GPT-5.5, and Gemini 3.5 Flash (the three commercial models via API, because the authors’ web-UI accounts are contaminated by conversation history about this illusion). n = 10 per cell, 480 trials in total.

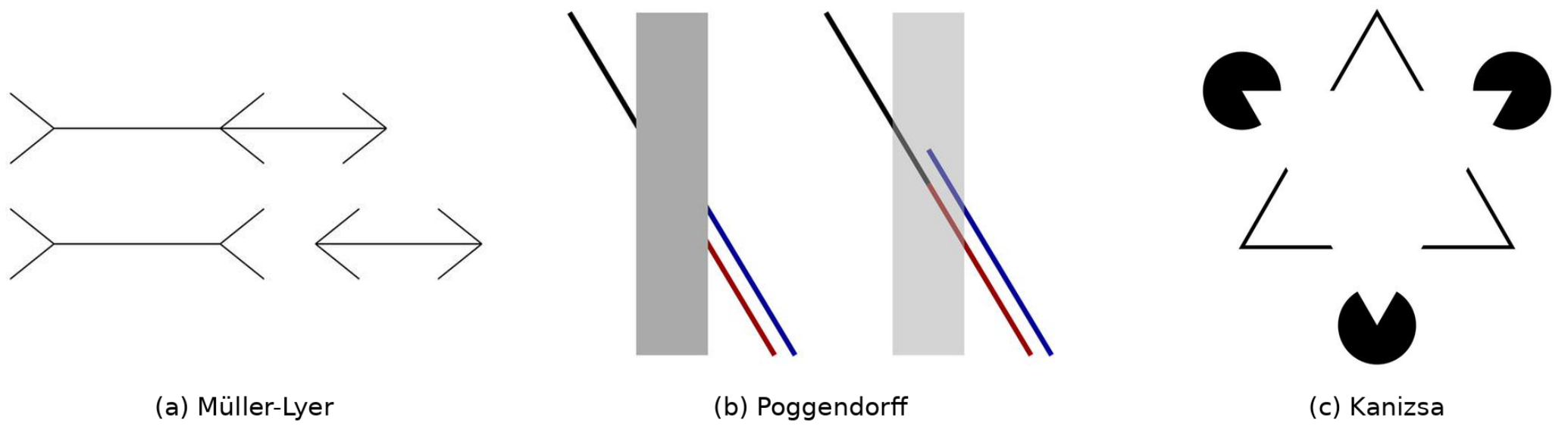

**Appendix 1—figure 1. Positive-control stimuli for the recognition probe.** (**a**) The Müller-Lyer illusion. (**b**) The Poggendorff illusion. (**c**) The Kanizsa illusion (illusory contours). All were presented under the same protocol as 104.jpg. Images are from Wikimedia Commons: (**a**) “Müller-Lyer Illusion - MathWorld version.svg” (public domain); (**b**) “Poggendorff illusion.svg” and (**c**) “Kanizsa triangle.svg” (both by Fibonacci, CC BY-SA 3.0).

**Main result (Appendix 1—table 1).**

**Appendix 1—table 1. Correct-naming rates for the naming questions (QV2 + QV3, n = 20 per cell).**

| Model | Watanabe | Müller-Lyer | Poggendorff | Kanizsa |
|---|---|---|---|---|
| Qwen2.5-VL-32B | 0/20 (0%) | 2/20 (10%) | 2/20 (10%) | 0/20 (0%) |
| Claude Opus 4.7 | 0/20 (0%) | 20/20 (100%) | 20/20 (100%) | 20/20 (100%) |
| GPT-5.5 | 0/20 (0%) | 20/20 (100%) | 20/20 (100%) | 20/20 (100%) |
| Gemini 3.5 Flash | 0/20 (0%) | 13/20 (65%) | 20/20 (100%) | 20/20 (100%) |

Claude Opus 4.7 and GPT-5.5 named all three classical illusions correctly 100% of the time. By contrast, the Watanabe Illusion was never named correctly in any of the 80 naming trials pooled across the four VLMs. Their performance on the classical illusions rules out the alternative that VLMs cannot name illusions in general, so this result is strong evidence that 104.jpg is unknown within the training distribution of today's major VLMs.

**Misnamings reflect structural proximity.** The Watanabe stimulus was misnamed as another illusion in 13 trials. The illusion names given in those trials (counted once per name when a trial gave several) were Poggendorff 8, Müller-Lyer 5, and Kanizsa 1. Poggendorff was the most frequent misnaming, and both illusions share the task structure of "mentally estimating the extension of a line segment whose visible extent is limited." VLMs extract the visual structure of the stimulus to some degree, but have not learned the particular name tied to that structure.

# Appendix 2. Statistical tables

**Appendix 2—table 1. Human reference data (n = 131).**

| Measure | Veridical | Mean (SD) | Median |
|---|---|---|---|
| $Q_1$ (from top) | 1 | 4.65 (2.63) | 4 |
| $Q_2$ (from bottom) | 11 | 6.36 (2.86) | 7 |
| $IIQ_1 = Q_1 - 1$ | 0 | 3.65 (2.63) | 3 |

| Measure | Veridical | Mean (SD) | Median |
|---|---|---|---|
| $IIQ_2 = 11 - Q_2$ | 0 | 4.64 (2.86) | 4 |
| $II$ | 0 | 4.15 (2.28) | 4.00 |
| Sum $Q_1 + Q_2$ | 12 | 11.01 (3.07) | 12 |

**Appendix 2—table 2. Distributional statistics for the six data sources.** All sources were processed with the same parser, the same directional normalization, and the sample standard deviation (ddof = 1). Because of extraction failure on some trials, the Illusion Index and the sum of the two answers use the trials in which both answers were extracted, paired by trial (paired n = 58 for Qwen-7B and 59 for Gemini).

| Measure | Human (n = 131) | Qwen-32B (n = 60) | Qwen-7B (n = 59; paired 58) | Sonnet 4.6 (n = 60) | GPT-5.5 Instant (n = 60) | Gemini 3.5 Flash (n = 59/60) |
|---|---|---|---|---|---|---|
| $Q_1$ — mean (SD) | 4.65 (2.63) | 5.22 (0.90) | 4.78 (2.51) | 4.30 (1.53) | 8.03 (1.34) | 7.10 (1.86) |
| $Q_2$ — mean (SD) | 6.36 (2.86) | 5.40 (0.83) | 4.86 (2.32) | 3.97 (1.73) | 2.48 (0.54) | 5.00 (1.52) |
| $IIQ_1$ — mean (SD) | 3.65 (2.63) | 4.22 (0.90) | 3.78 (2.51) | 3.30 (1.53) | 7.03 (1.34) | 6.10 (1.86) |
| $IIQ_2$ — mean (SD) | 4.64 (2.86) | 5.60 (0.83) | 6.14 (2.32) | 7.03 (1.73) | 8.52 (0.54) | 6.00 (1.52) |
| $II$ — mean (SD) | 4.15 (2.28) | 4.91 (0.63) | 4.95 (1.75) | 5.17 (1.13) | 7.78 (0.75) | 6.04 (1.29) |
| $II$ — range | [0.00, 9.50] | [3.50, 6.50] | [1.50, 8.50] | [2.00, 7.00] | [5.00, 9.00] | [2.50, 7.50] |

| Measure | Human (n = 131) | Qwen-32 B (n = 60) | Qwen-7B (n = 59; paired 58) | Sonnet 4.6 (n = 60) | GPT-5.5 Instant (n = 60) | Gemini 3.5 Flash (n = 59/60) |
|---|---|---|---|---|---|---|
| Sum $Q_1 + Q_2$ — mean (SD) | 11.01 (3.07) | 10.62 (1.19) | 9.66 (3.38) | 8.27 (2.36) | 10.52 (1.38) | 12.08 (2.23) |

**Appendix 2—table 3. Behavioral error vs. internal probe error (Qwen2.5-VL-32B).**

| Measure | Veridical | Behavioral response (mean) | Behavioral error | Internal probe error (n = 68) | Amplification |
|---|---|---|---|---|---|
| $Q_1$ (dot from top) | 1 | 5.22 | 4.22 dots | 0.71 dots | 5.9× |
| $Q_2$ (dot from bottom) | 11 | 5.40 | 5.60 dots | 0.71 dots | 7.9× |

**Appendix 2—table 4. Logit lens: final-layer digit-token probabilities (mean of n = 10 each).**

| Digit | v1 (“would hit the”) | v2 (“topmost dot is the”) | v3 (“number five is the”) | v4 (“would hit dot number”) |
|---|---:|---:|---:|---:|
| 1 | 0.039 | 0.944 | 0.029 | 0.090 |
| 4 | 0.177 | 0.004 | 0.020 | 0.190 |
| 5 | 0.292 | 0.007 | 0.858 | 0.190 |
| 6 | 0.201 | 0.007 | 0.014 | 0.168 |
| (others: 2, 3, 7–9) | ≤ 0.14 | ≤ 0.012 | ≤ 0.03 | ≤ 0.12 |

**Appendix 2—table 5. Prompt cancellation (n = 20 each, $Q_1$).**

| Prompt | Description | Mean | SD | Difference from P0 |
|---|---|---|---|---|
| P0 | Control | 5.45 | 0.89 | (reference) |
| P1 | Naming as "Watanabe Illusion" | 5.35 | 1.04 | −0.10 |
| P2 | Naming as "visual illusion" | 5.35 | 0.99 | −0.10 |
| P3 | Prior-knowledge suppression | 5.20 | 1.15 | −0.25 |
| P4 | Counter-hint | 4.20 | 1.01 | −1.25 |
| P5 | Forced visual chain of thought | 5.10 | 1.65 | −0.35 |
| P6 | Disclosure of the correct answer | 1.35 | 0.67 | −4.10 |
| P7 | Irrelevant context | 5.55 | 0.83 | +0.10 |

**Appendix 2—table 6. Semantic preamble intervention (PIE, one-shot, n ≈ 130 each, $Q_1$).**

| Condition | Category | Placement | $n_{valid}$ | Mean | SD | SD / human SD |
|---|---|---|---|---|---|---|
| Human | — | — | 131 | 4.65 | 2.63 | 1.00 |
| A | (baseline) | — | 129 | 5.52 | 0.74 | 0.28 |
| C | Character | user | 126 | 5.67 | 1.11 | 0.42 |
| E | Environment | user | 130 | 5.65 | 0.99 | 0.38 |
| G | Neutral | user | 130 | 5.52 | 0.94 | 0.36 |
| B | Character | system | 125 | 6.28 | 1.75 | 0.67 |
| D | Environment | system | 127 | 6.20 | 1.57 | 0.60 |
| F | Neutral | system | 128 | 6.40 | 1.55 | 0.59 |
| H | Char + Env | system | 122 | 6.36 | 1.66 | 0.63 |
| I | Char + Env | user | 127 | 5.83 | 1.22 | 0.46 |

# Appendix 3. Full text of the prompts used (Figure 4, Appendix 2—tables 5 and 6)

This appendix records, in reproducible form, the text actually given to the model in the prompt-cancellation experiment (Figure 4, Appendix 2—table 5) and the semantic preamble intervention experiment (PIE, Appendix 2—table 6). Both are single-turn trials on Qwen2.5-VL-32B-Instruct (A100) with no conversation history.

## The fixed part shared by all conditions

Both experiments share the following fixed preamble and fixed Q1 body (the same strings as Q1 in the n = 60 behavioral experiment).

**PREAMBLE_FIXED**

> Look at the image below straight on from the front. Please answer based on your visual impression alone — do not use any tools such as Python, image processing, a ruler, or a protractor to measure.

**Q1_BODY**

> Estimate which dot on the right edge the line in the circle would hit if extended. Answer like ‘the Xth dot from the top.’

**Appendix 3—table 1. Generation settings (common to both experiments).**

| Item | Value |
| --- | --- |
| model | Qwen/Qwen2.5-VL-32B-Instruct |
| do_sample | True |
| temperature | 1.0 |
| top_p | 1.0 |
| repetition_penalty | 1.05 |
| max_new_tokens | 512 |
| stimulus | 104.jpg (identical in every trial) |
| conversation history | none (each trial independent) |

## Prompt cancellation (Figure 4, Appendix 2—table 5)

Script run_qwen25vl_cancellation_a100_q1.py. Eight conditions × n = 20 = 160 trials, Q1 only. seed = (prompt index) × 1000 + trial_idx (P0 = 0–19, P1 = 1000–1019, and so on to P7 = 7000–7019).

The prompt is assembled as follows; the message consists of a single user turn (image + text).

```
final_prompt = PREAMBLE_FIXED
        + "\n\n"
        + <condition-specific body>
```

The <condition-specific body> is, except for P0, “the intervention sentence + a single space + Q1_BODY.”

**P0: control**

> Estimate which dot on the right edge the line in the circle would hit if extended. Answer like ‘the Xth dot from the top.’

**P1: name_specific (naming the particular illusion)**

> The image below shows the Watanabe Illusion. Estimate which dot on the right edge the line in the circle would hit if extended. Answer like ‘the Xth dot from the top.’

**P2: name_generic (naming the class of illusion)**

> The image below contains a visual illusion. Estimate which dot on the right edge the line in the circle would hit if extended. Answer like ‘the Xth dot from the top.’

**P3: knowledge_suppress (suppression of prior knowledge)**

> Ignore any prior knowledge about visual illusions or typical geometric figures. Answer based only on the line’s pixel-level geometry. Estimate which dot on the right edge the line in the circle would hit if extended. Answer like ‘the Xth dot from the top.’

**P4: counter_hint (a counter-hint toward the top)**

> For figures like this one, the line typically hits a dot near the top of the right edge. Estimate which dot on the right edge the line in the circle would hit if extended. Answer like ‘the Xth dot from the top.’

**P5: force_visual_cot (forced visual chain of thought)**

> First describe the line’s slope in your own words. Then Estimate which dot on the right edge the line in the circle would hit if extended. Answer like ‘the Xth dot from the top.’

**P6: ground_truth_disclose (disclosure of the correct answer)**

The correct answer is the 1st dot from the top. Independently, what is your own visual estimate? Estimate which dot on the right edge the line in the circle would hit if extended. Answer like ‘the Xth dot from the top.’

**P7: random_distractor (irrelevant context)**

While listening to instrumental music in a quiet library on a Tuesday afternoon, Estimate which dot on the right edge the line in the circle would hit if extended. Answer like ‘the Xth dot from the top.’

The capital letter in “Then Estimate” in P5 arises from string concatenation; the string is recorded here exactly as it was at run time.

## Semantic preamble intervention (PIE, Appendix 2—table 6)

The scripts are run_qwen25vl_pie_oneshot_a100.py (conditions A–G) and run_qwen25vl_pie_combined_a100.py (conditions H and I). The preamble set is pie_prompts_v2.csv (390 preambles = 3 categories × 130). The design is one-shot: one condition = 130 trials, each trial using a different preamble and a different seed. This mirrors the human data available at design time (n ≈ 130), in which each participant answered once; the final human analysis set (n = 131) is of almost the same size. No preamble is repeated.

### The question text (invariant across conditions)

Q1_USER_TEXT = PREAMBLE_FIXED + "\n\n" + Q1_BODY

That is, the question part of the user message is the two sentences above concatenated in that order, and it was never altered regardless of the presence or placement of a preamble.

### Conditions and message structure

**Appendix 3—table 2. PIE conditions and message structure.**

| Condition | Category | Placement | seed range | Message structure |
|---|---|---|---|---|
| A | (baseline) | — | 10000–10129 | user: [image, Q1_USER_TEXT] |
| B | Character | system | 11000–11129 | system: preamble / |

| Condition | Category | Placement | seed range | Message structure |
| --- | --- | --- | --- | --- |
| | | | | user: [image, Q1_USER_TEXT] |
| C | Character | user | 12000–12129 | user: [image, preamble + "\n\n" + Q1_USER_TEXT] |
| D | Environment | system | 13000–13129 | same form as B |
| E | Environment | user | 14000–14129 | same form as C |
| F | Neutral | system | 15000–15129 | same form as B |
| G | Neutral | user | 16000–16129 | same form as C |
| H | Character + Environment | system | 17000–17129 | system: concatenated preamble |
| I | Character + Environment | user | 18000–18129 | user: concatenated preamble + Q1_USER_TEXT |

The concatenation rule for H and I:

```
combined_preamble = <character preamble>
                  + "\n\n"
                  + <environment preamble>
```

Within each category, the 130 preambles are used in ascending order of (subcategory_code, numeric index). Because subcategory codes sort lexicographically, the actual order is C1-N, C1-S, C1-V, C1-X for Character; C2-P, C2-S, C2-V, C2-X for Environment; and C3-D, C3-E, C3-H, C3-S for Neutral. The pairing in H and I is a deterministic one-to-one correspondence based on this order.

**Appendix 3—table 3. Pairing of preambles in conditions H and I.**

| trial | Character | Environment |
|---|---|---|
| 0 | C1-N-01 (Lyric poet) | C2-P-01 (Zero gravity) |
| 1 | C1-N-02 (Tax accountant) | C2-P-02 (Underwater city) |
| … | … | … |
| 129 | C1-X-32 (Crystalline lifeform) | C2-X-32 (Inside a stained glass window) |

## Composition of the preamble set

**Appendix 3—table 4. Composition of the preamble set.**

| Category | Subcategory | ID prefix | n | First item (run order) |
|---|---|---|---|---|
| Character | neutral_profession | C1-N | 33 | Lyric poet |
| | spatial_cognition | C1-S | 32 | Architect |
| | vision_specialized | C1-V | 33 | Savanna hunter |
| | extreme_persona | C1-X | 32 | Thousand-year-old sage |
| Environment | physical_environment | C2-P | 32 | Zero gravity |
| | social_environment | C2-S | 33 | Quiet library |
| | visual_envi | C2-V | 33 | Dense fog |

| Category | Subcategory | ID prefix | n | First item (run order) |
|---|---|---|---|---|
| | ronment | | | |
| | extreme_environment | C2-X | 32 | Four-dimensional space |
| Neutral | daily_activity | C3-D | 33 | Sunday coffee |
| | emotion_mood | C3-E | 32 | Calm satisfaction |
| | hobby_activity | C3-H | 32 | Balcony gardening |
| | static_situation | C3-S | 33 | Waiting for a train |

Each category consists of four subcategories totaling 130 preambles (33 + 32 + 33 + 32). Preamble length is 16–38 words (mean 26.7). IDs run from 01 within each subcategory, with no duplicates and no gaps. The Neutral category deliberately excludes spatial and visual vocabulary. The first design carried 25 preambles per subcategory, 300 in total; to obtain a one-to-one correspondence with the human n ≈ 130 available at design time, 7–8 were added to each subcategory (ID 26 onward, 90 in total), giving 390. The 300 from the first version are unchanged in both ID and wording.

## Example preambles (one from each of the 12 subcategories)

**Appendix 3—table 5. Example preambles.**

| ID | Full text |
|---|---|
| C1-N-01 | You are a lyric poet who spends most days searching for the precise word to describe a feeling. Numbers and measurements rarely enter your thoughts; what matters is rhythm and image. |

| ID | Full text |
|---|---|
| C1-S-01 | You are an architect with three decades of experience designing complex buildings. You instinctively read lines, angles, and proportions in everything you see, and your mind constantly reconstructs spaces in three dimensions. |
| C1-V-01 | You are a hunter from a nomadic tribe in the African savanna. Your eyes are trained to spot prey at extreme distances, and you can see fine details that most people would miss in the shimmering heat haze. |
| C1-X-01 | You are a sage who has lived for a thousand years on a remote mountain. You have watched dynasties rise and fall, and you perceive the world with a patience that no mortal could possess. |
| C2-P-01 | You live in a world without gravity. Objects drift freely in every direction, and your sense of "up" and "down" has dissolved into a pure three-dimensional awareness of space. |
| C2-S-01 | You spend your days in the silent reading room of an ancient library. The only sounds are the rustle of paper and the distant ticking of a clock, and your attention has become slow and careful. |
| C2-V-01 | You live in a forest where dense fog never lifts. You have learned to see the world through soft, diffuse light, where edges blur and distances are estimated by sound and intuition |

| ID | Full text |
| --- | --- |
| | rather than by sight. |
| C2-X-01 | You live in a four-dimensional space. What three-dimensional beings call “lines” and “planes” are merely shadows of the richer structures you perceive natively. |
| C3-D-01 | You are sipping a warm cup of coffee on a quiet Sunday morning. The newspaper is folded beside you, and soft sunlight gently fills the kitchen. |
| C3-E-01 | You feel a quiet sense of satisfaction. Nothing dramatic has happened today, but several small things went well, and you find yourself unusually at peace. |
| C3-H-01 | You are tending to a small herb garden on your balcony. You have just finished watering the basil and are pinching back the mint to encourage new growth. |
| C3-S-01 | You are standing on a station platform, waiting for a train that is six minutes away. A mild breeze moves through the station, and you have nothing in particular to do. |

The full text of all 390 preambles is released in the repository as pie_prompts_v2.csv.

## A completed prompt (condition I, trial 0: C1-N-01 + C2-P-01)

You are a lyric poet who spends most days searching for the precise word to describe a feeling. Numbers and measurements rarely enter your thoughts; what matters is rhythm and image.

You live in a world without gravity. Objects drift freely
in every direction, and your sense of "up" and "down" has
dissolved into a pure three-dimensional awareness of space.

Look at the image below straight on from the front. Please
answer based on your visual impression alone — do not use
any tools such as Python, image processing, a ruler, or a
protractor to measure.

Estimate which dot on the right edge the line in the circle
would hit if extended. Answer like 'the Xth dot from the
top.'

## Notes

- Both experiments target Q1 (from top) only; Q2 was not run.
- As a remnant of the first design, pie_prompts_v2.csv contains a full_prompt column (all 390 rows using the connective “Now,”), but the execution scripts do not read it. The scripts read only the preamble column and use the Q1_USER_TEXT above for the question part. The wording in the full_prompt column is therefore not the string used in this experiment.

# Appendix 4. Reproduction map of code and data

All code, stimuli, and raw data are available at GitHub (https://github.com/eijwat/watanabe-illusion-vlm) and archived at Zenodo (https://doi.org/10.5281/zenodo.23096369). The repository is organized as follows.

```
watanabe-illusion-vlm/
  stimuli/          stimulus images and ground truth
  scripts/          experiment and analysis scripts
  prompts/          prompt definitions and protocols
  data/human/       cleaned responses of the human experiment (Gorilla)
  data/manual_logs/ manual logs of commercial models
  results/          raw output of each experiment
  requirements.txt
```

Below we list, experiment by experiment, the programs and data required to reproduce every result in this paper.

## Stimuli and experimental materials

| File | Contents |
|---|---|
| stimuli/104.jpg | Main stimulus (539 × 264 px, segment 23.5°, 11 dots on the right edge, MD5 9cba67ee1a9ccb213dcab20206e4135e) |
| scripts/generate_stimuli.py | Parametric stimulus generation (17 angles × 4 conditions = 68 images) |
| stimuli/stimulus_metadata.csv | Ground truth for the 68 images (angle, intersection coordinates, dot count, circle parameters) |
| scripts/measure_stimulus_v2.py | Verification script measuring segment angle, dot positions, and the correct dot directly from the image |
| scripts/make_fig1.py – make_fig4.py, make_figA1.py | Drawing of Figures 1–4 and Appendix 1—figure 1 (correspondence below) |
| stimuli/control_muller_lyer.png, control_poggendorff.png, control_kanizsa.png | Classical-illusion positive controls (Appendix 1—figure 1; from Wikimedia Commons, with sources and licenses recorded in the repository README) |
| stimuli/G1.png, G2.png, G3.png | The screens as actually presented in the human experiment (Gorilla) |
| stimuli/104_jpg_ChatGPT_Image.png | Output image of the generative probe (Figure 1b) |

The 68 parametric stimuli can be regenerated from generate_stimuli.py and stimulus_metadata.csv.

## Human experiment (Figure 1c–e, Appendix 2—table 1)

| File | Contents |
|---|---|
| data/human/human_1st_test.csv, human_2nd_test.csv | Cleaned human responses ($Q_1$, $Q_2$, $Q_3$) |

| File | Contents |
| --- | --- |
| data/human/human_2nd_test_Q1Q2.csv | $Q_1$/$Q_2$ complete-case extract |

The human reference data used in the body is human_2nd_test.csv; the n = 131 cases in which both $Q_1$ and $Q_2$ were answered within the range [1, 11] are the analysis set. The raw Gorilla exports are not released because they contain participant-identifying information.

## VLM behavioral experiment (Figure 1c–e, Appendix 2—tables 1 and 2)

Open-weight models:

| File | Contents |
| --- | --- |
| scripts/run_qwen25vl_n60_a100_q1q2.py | n = 60 behavioral experiment for Qwen2.5-VL-32B / 7B (A100) |
| scripts/run_qwen25vl_n60_spark_q1q2.py | The same, DGX Spark version |
| results/qwen25vl_n60_a100_q1q2/ | Raw output for 32B (n60_progress.json, input_verification_meta.json) |
| results/qwen25vl_n60_a100_q1q2_7b/ | Raw output for 7B (same files) |

Commercial models (manual web-UI runs):

| File | Contents |
| --- | --- |
| prompts/commercial_vlm_protocol.md | Protocol for the commercial web-UI experiment |
| prompts/prompts.txt | Manual run procedure and full text of the presented prompts |
| data/manual_logs/ manual_log_template_n60.md | Full responses of the three commercial models (canonical data) |
| scripts/parse_manual_log.py | Manual log to JSON conversion |

Integrated analysis:

| File | Contents |
|---|---|
| scripts/analyze_commercial_vlm.py | **Canonical analysis pipeline.** Integrates the six sources with one parser, one directional normalization and ddof = 1, and outputs the integrated statistics (stats_extended.json, table2_extended.md) and the material for Figure 1c–e. Source of the numbers in Appendix 2—tables 1 and 2 |
| scripts/plot_vlm_vs_human.py, plot_qwen25vl_n60_v2_vs_human.py | Single-model versus human comparison figures |
| scripts/analyze_n60_q1q2.py, analyze_bias.py, analyze_qwen25vl_n60_v2.py | Auxiliary analyses |

The outputs of the analysis (statistics JSON, tables, figures) are not included in the repository; they are produced by running the scripts above against the raw output.

The reproduction procedure is as follows. For the two Qwen sizes, run run_qwen25vl_n60_a100_q1q2.py. For the three commercial models, record the web-UI responses in a manual log following commercial_vlm_protocol.md and prompts.txt, then convert to JSON with parse_manual_log.py. Passing the five JSON files and human_2nd_test.csv to analyze_commercial_vlm.py then produces Appendix 2—tables 1 and 2 and the material for Figure 1c–e. The response JSON files of the three commercial models (n60_progress_sonnet46.json, n60_progress_gpt55.json, n60_progress_gemini35flash.json) are derivatives of the manual logs and are therefore not included in the repository; they are generated from the manual logs by parse_manual_log.py.

## Vision encoder probe (Figure 2, Appendix 2—table 3)

| File | Contents |
|---|---|
| scripts/analyze_qwen25vl_vit_a100.py | Ridge linear probes on ViT / Patch Merger output (leave-one-out) |
| scripts/check_vit_letterbox_shape.py | Verification of the input preprocessing |

| File | Contents |
| --- | --- |
| | shape |
| scripts/analyze_vision_encoders.py | Analysis across several encoders |
| scripts/decoder_model.py, train_decoder.py | Decoder-type probe (auxiliary) |
| results/qwen25vl_vit_probe_a100/ probe_results.csv, probe_results.json | $R^2$ and MAE per probe × condition (source of the numbers in Figure 2 and Appendix 2—table 3) |

Running this requires the 68 images produced by generate_stimuli.py together with stimulus_metadata.csv. The ViT / Patch Merger representation cache representations.npz (about 567 MB) is not included in the repository because of its size; it is regenerated by running the same script.

## Logit lens (Figure 3, Appendix 2—table 4)

| File | Contents |
| --- | --- |
| scripts/run_qwen25vl_logit_lens_a100.py (v1) | continuation “The line would hit the” |
| scripts/ run_qwen25vl_logit_lens_v2_a100.py | continuation “The topmost dot is the” |
| scripts/ run_qwen25vl_logit_lens_v3_a100.py | continuation “The number five is the” |
| scripts/ run_qwen25vl_logit_lens_v4_a100.py | continuation “The line would hit dot number” |
| scripts/plot_logit_lens_4panel.py | Draws the layer × digit-probability curves for the four conditions |
| scripts/analyze_intermediate_layers.py | Intermediate-layer analysis (auxiliary) |
| results/qwen25vl_logit_lens_v{1..4}_a100/ layer_logits.npz, layer_logits_meta.json | Layer-wise logits and probabilities (source of the numbers in Figure 3 and Appendix 2—table 4) |

For each continuation, n = 10 trials, reading the within-layer softmax probabilities of digit tokens 1–9 at the final token position across the 64 layers of the language model.

## Prompt cancellation (Figure 4, Appendix 2—table 5)

| File | Contents |
|---|---|
| scripts/ run_qwen25vl_cancellation_a100_q1.py | The eight conditions × n = 20 run (full prompt text in Appendix 3) |
| scripts/plot_cancellation_v1.py | Drawing and computation of the condition-wise statistics |
| results/qwen25vl_cancellation_v1_a100/ | Full responses of the 160 trials (cancellation_progress.json). The condition-wise mean / SD (cancellation_stats.json, source of the numbers in Appendix 2—table 5) is generated from this file by plot_cancellation_v1.py |

## Semantic preamble intervention (PIE, Appendix 2—table 6)

| File | Contents |
|---|---|
| prompts/pie_prompts_v2.csv | The 390 preambles (Appendix 3) |
| prompts/pie_prompts_v1.md | Preamble design document (first version, 300 preambles) |
| scripts/ run_qwen25vl_pie_oneshot_a100.py | Conditions A–G (n = 130 each, 910 trials) |
| scripts/ run_qwen25vl_pie_combined_a100.py | Conditions H and I (n = 130 each, 260 trials) |
| results/pie_oneshot_20260527_005108/ | Raw output for conditions A–G (config.json, responses.jsonl, condition_summary.csv) |
| results/pie_combined_20260527_043814/ | Raw output for conditions H and I (same files) |

## Recognition probe (Appendix 1, Appendix 1—table 1)

| File | Contents |
| --- | --- |
| scripts/ run_qwen25vl_recognition_stage2_a100.py | Recognition probe for Qwen2.5-VL-32B (4 stimuli × 3 questions × n = 10) |
| scripts/ run_commercial_recognition_stage2.py | API runs for the three commercial models (Claude Opus 4.7, GPT-5.5, Gemini 3.5 Flash) |
| scripts/analyze_recognition_stage2.py | Aggregation of naming rates, yes/no, and the misnaming breakdown |
| results/ qwen25vl_recognition_stage2_a100/ | Full responses of Qwen (progress.json, responses.jsonl, summary.csv) |
| results/commercial_recognition_stage2/ | Full responses and aggregate CSV of the three commercial models (source of the numbers in Appendix 1—table 1) |

Running the commercial APIs requires the environment variables ANTHROPIC_API_KEY, OPENAI_API_KEY, and GOOGLE_API_KEY (or GEMINI_API_KEY). The total number of API calls is 360.

## Correspondence between the figures and their sources

| Figure | Drawing script | Input |
| --- | --- | --- |
| Figure 1 | scripts/make_fig1.py | stimuli/104.jpg, stimuli/104_jpg_ChatGPT_Image.png, stats_extended.json from analyze_commercial_vlm.py |
| Figure 2 | scripts/make_fig2.py | fig_angle_probe.png and fig_spatial_probe.png from analyze_qwen25vl_vit_a100.py |

| Figure | Drawing script | Input |
| --- | --- | --- |
| Figure 3 | scripts/make_fig3.py | results/ qwen25vl_logit_lens_v{1..4} _a100/layer_logits.npz and layer_logits_meta.json |
| Figure 4 | scripts/make_fig4.py | cancellation_stats.json output by plot_cancellation_v1.py |
| Appendix 1—figure 1 | scripts/make_figA1.py | stimuli/ control_muller_lyer.png, control_poggendorff.png, control_kanizsa.png |

None of the drawing scripts writes a figure number or title into the image; the legend carries the explanation. Colors are assigned consistently across all figures, with crimson (#dc143c) for the geometric ground truth and pale cyan (#e5ffff) for the reference band (the middle-layer band in Figure 3, the P0 ± 1 SD band in Figure 4).

Figures 1, 3, and 4 are drawn directly from numerical data. Figure 2 alone is composed by taking panels out of the output figures of the angle probe and the spatial probe; given a prediction-value output from analyze_qwen25vl_vit_a100.py it could likewise be replaced by direct drawing.

## Execution environment

| Item | Contents |
| --- | --- |
| Open-weight run machine 1 | NVIDIA A100 80GB (x86_64), Docker image watanabe-illusion:a100 |
| Open-weight run machine 2 | NVIDIA DGX Spark (GB10 Grace Blackwell, 128GB unified memory, ARM64, CUDA 13.0), Docker image watanabe-illusion:phase5 |
| Models | Qwen/Qwen2.5-VL-32B-Instruct, Qwen/Qwen2.5-VL-7B-Instruct (BF16, native dynamic resolution) |

| Item | Contents |
| --- | --- |
| Generation settings (behavioral and intervention experiments) | temperature = 1.0, top_p = 1.0, repetition_penalty = 1.05, max_new_tokens = 512, fixed per-trial seed |
| Commercial models (behavioral experiment) | Web UI default mode, memory storage off, each trial in a new temporary chat (as of May 2026) |
| Commercial models (recognition probe) | Official API, no conversation history |
| Human experiment | Gorilla online experiment platform |
| Statistics | Sample standard deviation (ddof = 1); directional normalization maps value to 11 − value + 1 |
| Dependencies | requirements.txt (A100 / Spark) |